\documentclass[fleqn, 10pt]{wlscirep}
\usepackage[squaren]{SIunits}
\title{Luminescence engineering in plasmonic meta-surfaces}

\author[1,2,3]{Tapashree Roy}
\author[1,4]{Edward T F Rogers}
\author[1,5*]{Nikolay I Zheludev}

\affil[1]{Optoelectronics Research Centre and Centre for Photonic Metamaterials, University of Southampton, Highfield, Southampton, SO17 1BJ, UK}
\affil[2]{Center for Nanoscale Materials, Argonne National Laboratories, Lemont, Illinois 60439, USA}
\affil[3]{School of Engineering and Applied Sciences, Harvard University, Cambridge, Massachusetts 02138, USA}
\affil[4]{Institute for Life Sciences, University of Southampton, Highfield, Southampton, SO17 1BJ, UK}
\affil[5]{Centre for Disruptive Photonic Technologies, Nanyang Technological University, Singapore 637371}

\affil[*]{niz@soton.ac.uk}

\begin{abstract}
Photoluminescence is a phenomenon of significant interest due to its wide range of technological applications in plasmonics, nanolasers, spasers, lasing spasers, loss compensation and gain in metamaterials, and luminescent media. Nanostructured materials are known to have very different luminescence characteristics to bulk samples or planar films. Here we show that by engineering a nanostructured meta-surface, we can choose the position of photoluminescence absorption and emission lines of thin gold films. The nanostructuring also aids to strong enhancement of the emission from gold --- by a factor of 76 in our experiments. This enhancement is determined by the relative position of the engineered absorption and emission lines to the exciting laser wavelength and the intrinsic properties of the constituent material. These luminescence-engineered materials combined with a resonant material, as in the lasing spaser, or with the power of reconfigurable metamaterials promise huge potential as tunable nanoscale light sources. 
\end{abstract}
\begin{document}

\flushbottom
\maketitle

\thispagestyle{empty}

\section*{Introduction}

One of the possible outcomes of light-matter interaction is emission of light from a matter due to radiative recombination of photo-excited electrons with holes, or photoluminescence. The first observation of photoluminescence was reported as early as 1845 by Sir John Herschel~\cite{Herschel1845,Lakowicz2006} and in 1928, Wood and Gaviola~\cite{Wood1928} observed that a two-step absorption with a real intermediate state gave a photoluminescence intensity that scaled with the square of the excitation intensity. In 1931 Maria Goeppert-Mayer predicted the existence of the two-photon absorption process~\cite{Goeppert-Mayer1931} where the ground state atoms absorb simultaneously two photons through a virtual intermediate state. It was not, however, until 30 years later that two-photon absorption was experimentally verified by Kaiser and Garret~\cite{Kaiser1961} after the first functioning laser became available~\cite{Maiman1960}. Two-photon absorption only takes place when there is a high probability of two photons being in the same place at the same time~\cite{Lakowicz2006}; hence two-photon excitation needs much higher incident intensities than the single photon case. The probability of two-photon absorption, as originally calculated by Maria Goeppert-Mayer~\cite{Goeppert-Mayer1931}, is proportional to the square of the incident intensity; whereas the probability of single photon absorption is proportional to the incident intensity.

In 1969 Mooradian reported the first observation of photoluminescence from metals such as gold and copper, as well as gold-copper alloys~\cite{Mooradian1969}. For gold at room temperature and illuminated with a \unit{488}{\nano\metre} continuous wave laser, a luminescence peak at \unit{540}{\nano\metre} was reported. Mooradian proposed that the luminescence resulted from direct recombination of $sp$ conduction band electrons just below the Fermi level ($\textrm{E}_F$) with the upper $d$ band holes. In 1986 Boyd \emph{et al.}~\cite{Boyd1986} reported new features in the emission spectra of the metals used by Mooradian for different excitation energies, which they attributed to different interband transitions. They also reported increased luminescence on roughened surfaces of metals due to localised surface plasmons. Since then surface plasmon enhanced photoluminescence from gold has been used to study various light-matter interactions and physical phenomena. Infrared luminescence from intraband transitions in gold was studied using rough gold films~\cite{Beversluis2003}. The field distributions of gold nanoparticles were studied by observing surface plasmon enhanced two-photon luminescence from the structures~\cite{Bouhelier2003}. Presently a large body of work exists on the effect of surface plasmons on gold nanoparticles and nanorods and on the nature of photoluminescence from such systems~\cite{Link2003,Imura2006,Wang2009,Beermann2012,Jiang2013,Tao2013,Wang2013,Zhao2012}.

\begin{figure} [ht]
\centering
\includegraphics[width=0.85\linewidth]{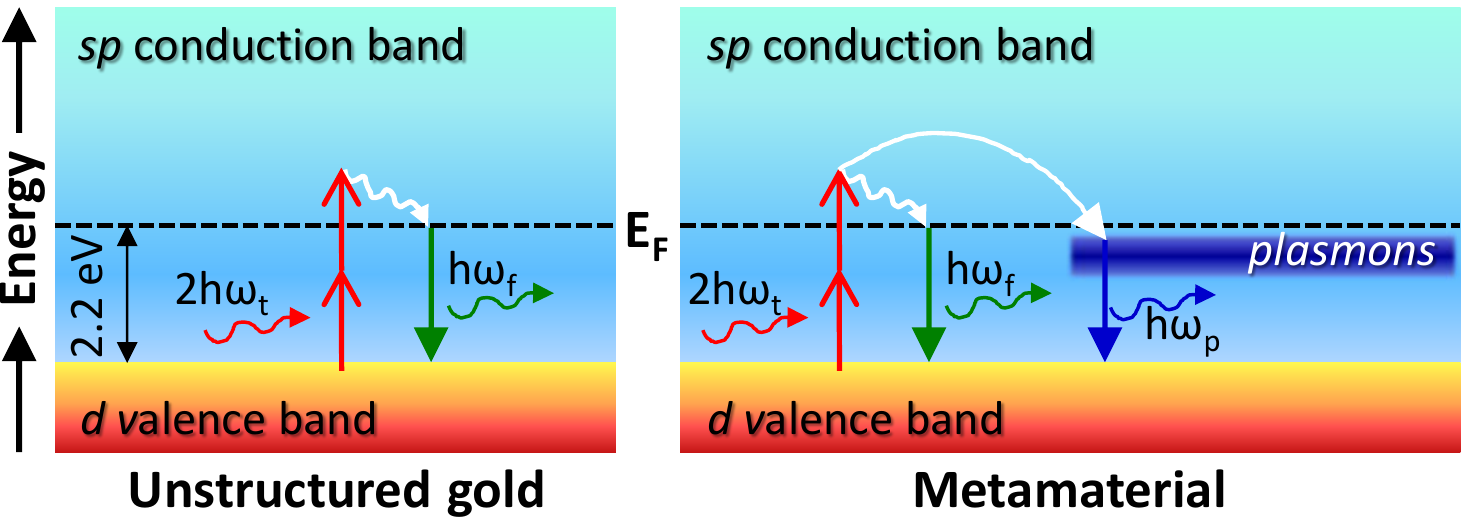}
\caption{Jablonski diagram explaining two-photon luminescence from unstructured gold and metamaterials.}
\label{fig-jab}
\end{figure}

In this paper, we experimentally study two-photon photoluminescence from an ultra-thin film of gold and compare with the photoluminescence from nanostructured gold films or, metamaterials. The emission peak of the continuous gold film is measured at \unit{535}{\nano\metre} which agrees with the original studies by Mooradian~\cite{Mooradian1969} and Boyd~\cite{Boyd1986}. In the presence of the metamaterials, a shift in the emission peak and a significant enhancement in luminescence intensity are observed compared with unstructured gold film. The photoluminescence emission peaks from the different metamaterials can be linked to the respective frequencies of absorption resonances. We report that resonant excitation at the absorption resonance of a metamaterial enhances photoluminescence by a factor of more than 76.

Enhancement of photoluminescence in the proximity of noble metal nanoparticles due to strongly localised surface plasmons has been reported in several other media, such as rare-earth ions~\cite{Rivera2012}, and cadmium sulphide nanoparticles~\cite{Nahm2013}. Surface plasmon enhanced photoluminescence has been studied in silver nanowires~\cite{Song2012} and gold nanoparticles and nano-sea-urchins~\cite{Su2010}. Strong photoluminescence due to two-photon excitation of roughened surfaces of noble metals was studied almost 30 years ago by Boyd \emph{et al.}~\cite{Boyd1986}. This enhancement in photoluminescence from rough surfaces under multi-photon excitation had been attributed to local-field enhancement due to surface plasmons. On the other hand, planar metamaterials are known to support strongly localised surface plasmons and the nature of plasmonic excitation has been studied~\cite{Rockstuhl2006}. Planar metamaterials using the asymmetric split ring unit-cell studied here, have also been reported to enhance nonlinearity of quantum dots~\cite{Tanaka2010}, graphene~\cite{Nikolaenko2012}, the planar gold film itself~\cite{Ren2011}, as well as enhance luminescence from carbon nanotubes~\cite{Nikolaenko2010}. These engineered plasmonic meta-surfaces are known to become luminescent on being excited by a highly loaclized free-electron beam~\cite{Adamo2012} and when combined with a resonant material, as in the lasing spaser~\cite{Plum2009}, or with the power of reconfigurable metamaterials~\cite{Adamo2013} they have known to offer huge potential as tunable nanoscale light sources. In this paper, we use these two dimensional metamaterials to study the enhancement of photoluminescence from the constituent gold films under two photon excitation.

Figure~\ref{fig-jab} depicts the process of two-photon luminescence from continuous and structured gold. Two photons are absorbed simultaneously exciting the electrons across the Fermi energy of gold (\unit{2.2}{\electronvolt}~\cite{Mooradian1969}) and the energised electrons in the $sp$ band recombine radiatively with holes in the $d$ band. When the gold film is structured into a metamaterial, localised surface plasmons are introduced. These plasmonic resonances increase the density of electronic states over certain energy levels which may couple with those of gold, giving different emission energies than those of the continuous film.

\section*{Results}

\subsection*{Luminescence from unstructured gold film}
\begin{figure}[ht]
\centering
\includegraphics[width=0.85\linewidth]{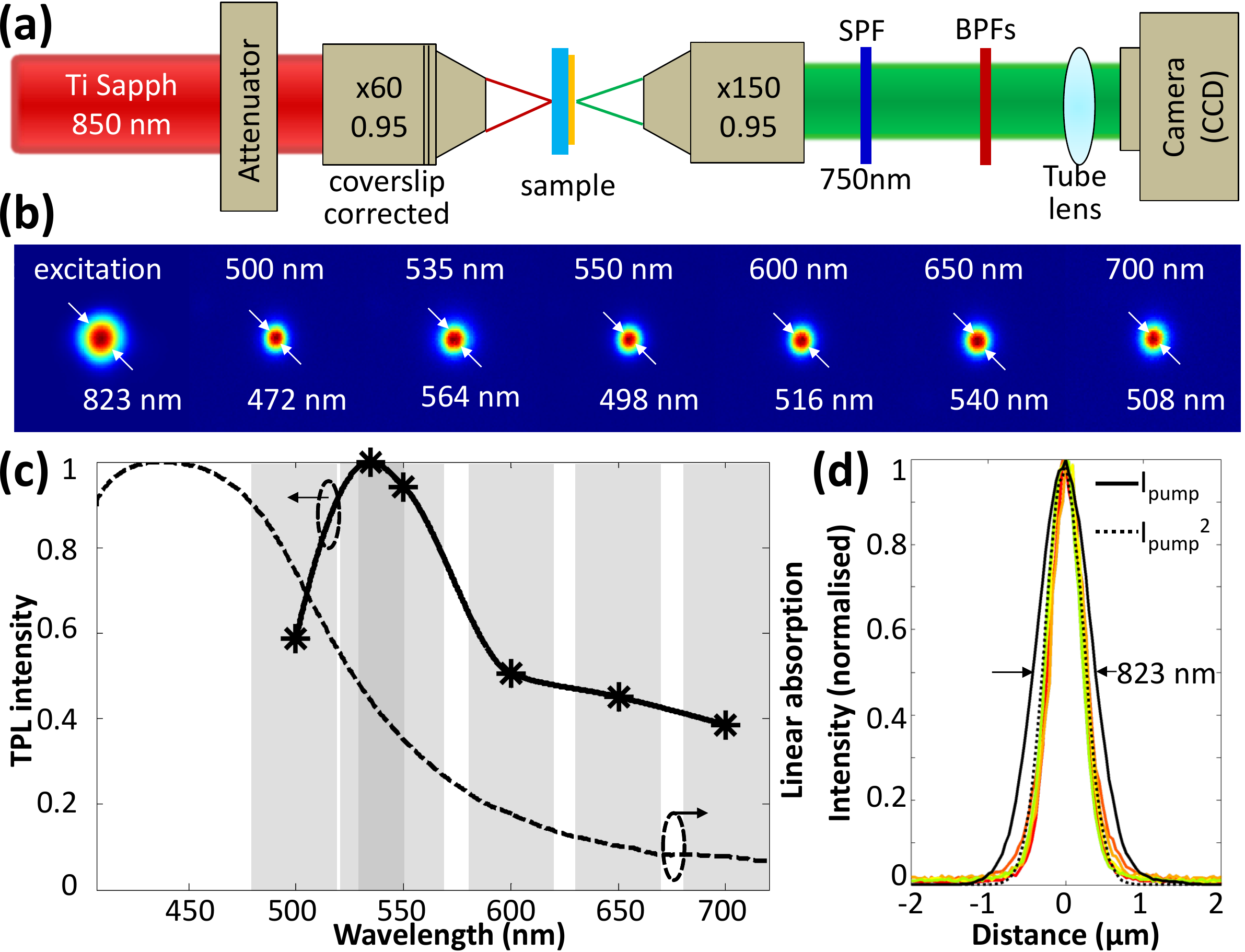}
\caption{Two-photon luminescence from gold film (a) Experimental set-up for measuring two-photon luminescence. SPF: short pass filter, BPF: band pass filter. (b) Normalised intensity images of the excitation spot, and the luminescence detected with different spectral filters (centre wavelength shown at the top). (c) Linear absorption and two-photon emission spectra. (d) Line-out through each frame in (b).}
\label{fig-au-tp}
\end{figure}

We begin by studying two-photon luminescence from a plain gold film. The sample is prepared by thermally depositing a \unit{50}{\nano\metre} thin film of gold on a \unit{170}{\micro\metre} thick (SiO$_{2}$) substrate. A schematic representation of the experimental set-up used for measuring the spectrally and spatially resolved photoluminescence is shown in Fig.~\ref{fig-au-tp}a. Ultra-short pulses (wavelength \unit{850}{\nano\metre}, duration \unit{75}{\femto\second}, repetition rate \unit{80}{\mega\hertz}) from a Ti-Sapphire laser (Chameleon Vision-S) are tightly focused with a microscope objective (NA=0.95, 60x, coverslip corrected) into the spot shown in Fig.~\ref{fig-au-tp}b (frame 1). An average power of \unit{31}{\milli\watt}, i.e., peak power density of $\sim\unit{0.98}{\tera\watt}/{\centi\metre}^2$ is incident on the sample. The photoluminescence is detected in transmission and imaged with a high numerical aperture objective (NA=0.95, 60x). The luminescence is spectrally resolved by a series of band-pass filters over different wavelengths ($\unit{500}{\nano\metre}$ to $\unit{700}{\nano\metre}$). Additionally a short-pass filter (cut off \unit{750}{\nano\metre}) is used to block the excitation beam. Due to the chromatic aberration of the lenses used, the light coming through each set of spectral filters may focus in a different plane. We correct this by refocussing the imaging objective for each case. A low-noise sCMOS camera (16-bit resolution 5 megapixel Andor Neo sCMOS) is used to record the the photoluminescence. The in-built GVD pre-compensator of the laser~\cite{Coherent2012} is used to correct for dispersion introduced by the optical components in the system. The pre-chirp applied was adjusted to maximise two-photon luminescence in one experiment and fixed for all subsequent experiments.

Figure~\ref{fig-au-tp}c shows the linear absorption and the normalised two-photon emission spectra from the ultra-thin gold film. The linear absorption of the gold film is calculated from the measured values of transmittance and reflectance obtained using a commercial spectrophotometer (Craic Technologies). For the emission spectra, each grey band depicts the bandwidth of the spectral filter used for detection of the luminescence. An emission peak is detected around \unit{535}{\nano\metre} which agrees with previous reports on photoluminescence from gold~\cite{Mooradian1969,Boyd1986,Beversluis2003}. It may be noted that the luminescence peak is red-shifted with respect to the linear absorption peak. This is attributed to the Stoke's shift~\cite{Stokes1852,Lakowicz2006}, which arises mostly due to non-radiative energy loss in the excited state before the emission process. Further, the excited electron may decay to higher vibrational level of the ground state accounting for lower energy (higher wavelength) of emission compared to absorption.

Figure~\ref{fig-au-tp}b shows the intensity images of the pump beam (no filter in the detection path), and the luminescence from the gold film at different emission wavelengths. The luminescence spots are smaller than the excitation spot size. This is better depicted in Fig.~\ref{fig-au-tp}d where the line-outs through each of the frames of Fig.~\ref{fig-au-tp}b are plotted. To verify that the luminescence spots result from the two-photon excitation of the atoms, the square of the excitation beam (dotted black) is also plotted in Fig.~\ref{fig-au-tp}d and is seen to be the same shape as the luminescence spots, demonstrating a quadratic dependence of luminescence on the incident intensity.

\subsection*{Luminescence from engineered metamaterials}

Next we study the two-photon luminescence from several planar metamaterials and compare with the unstructured gold film. The metamaterials are designed to engineer localised surface plasmon resonances on the ultra-thin gold film. As briefly discussed above, the two-photon luminescence on such structured gold films is a coupled effect of two-photon luminescence of the bulk gold as well as the surface plasmon-assisted local-field enhancement at both the emission and excitation wavelengths. This may result in a spectral shift and often a large enhancement of luminescence intensity due to the metamaterials.

\begin{figure}
\centering
\includegraphics[height=0.85\textheight]{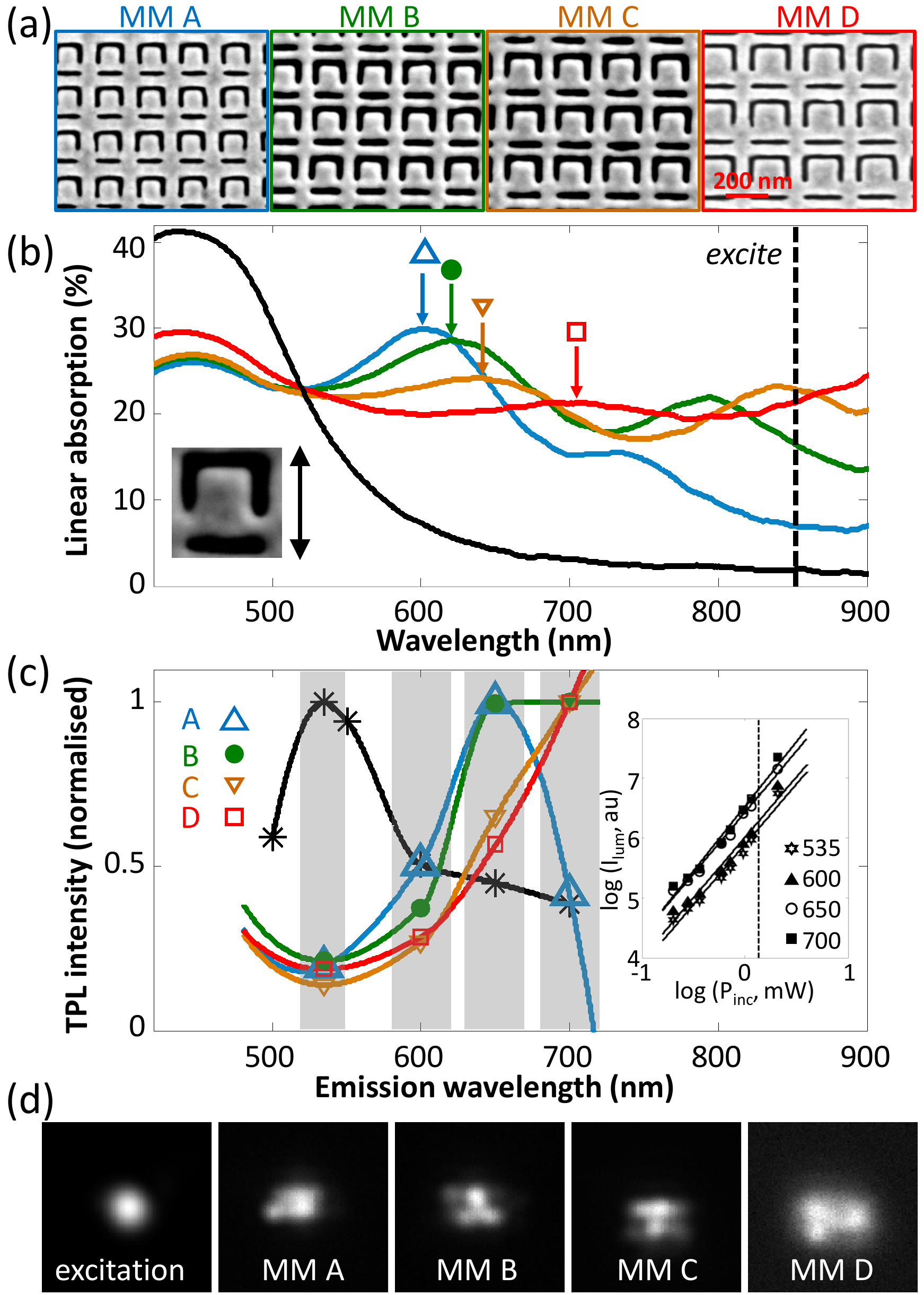}
\caption{Two-photon luminescence from metamaterials (a) Scanning electron micrographs of the different metamaterials. (b) Linear absorption spectra of plain gold film and the metamaterials. The vertical dashed line shows the excitation wavelength. Inset: incident polarisation with respect to the metamolecule. (c) Two-photon emission spectra of metamaterials and plain gold film. Inset: power dependence of metamaterial C for different detection spectral range. (d) Excitation spot on continuous gold film and luminescence from the different metamaterials detected at $\unit{700}\pm\unit{20}{\nano\metre}$.}
\label{fig-mm-tp}
\end{figure}

Figure~\ref{fig-mm-tp} summarises the linear absorption and two-photon luminescence spectra of the planar metamaterials compared to unstructured gold. Four different metamaterials (Fig.~\ref{fig-mm-tp}a) with the smallest metamolecule size of \unit{210}{\nano\metre} and increasing in \unit{20}{\nano\metre} steps are fabricated by focused-ion-beam milling of \unit{50}{\nano\metre} thin gold film. Each metamaterial contains approximately 8000 metamolecules. As with unstructured gold, the linear absorption spectra of the metamaterials is calculated from the transmittance and reflectance measured by the spectrophotometer. As shown in Fig.~\ref{fig-mm-tp}b the absorption spectra of the plain film shows a single peak around \unit{450}{\nano\metre}. However, nano-structuring of the same gold film introduces additional resonant peaks in the red/near-IR spectral range. The absorption resonance peaks red-shift with increasing metamolecule size.

As in the case of a continuous gold film, the metamaterials are excited with tightly focused, ultra-short pulses of wavelength \unit{850}{\nano\metre}. It must be noted that for the excitation of the unstructured gold film an average power of $\sim\unit{31}{\milli\watt}$ was used. In a systematic study, we found that the luminescence from the gold film falls below the noise level of the detector for incident average power less than $\sim\unit{10}{\milli\watt}$. However for the metamaterials an incident average power of $\sim\unit{1}{\milli\watt}$ is sufficient to record bright luminescence. The damage threshold for the metamaterials is only $\sim\unit{3}{\milli\watt}$. Hence, we choose an average incident power of $\sim\unit{1.35}{\milli\watt}$, i.e., a peak power density of $\unit{40}{\giga\watt}/{\centi\metre}^2$ to record the emission spectra from the metamaterials. This power density is almost 25 times lower than that used for recording luminescence from the continuous gold film.

Figure~\ref{fig-mm-tp}c compares the emission spectra of the metamaterials to plain gold film. The metamaterial A with the smallest unit-cell shows a clear emission peak within the spectral range of $\unit{650}\pm\unit{20}{\nano\metre}$. For the other three metamaterials a definite emission peak could not be inferred due to the low spectral resolution of the detection system. However, looking at the trend of the emission spectra it may be speculated that the emission peak of the metamaterial B lies somewhere in between \unit{650}{\nano\metre} and \unit{700}{\nano\metre}, while for the metamaterials C and D, the emission peaks lie within or beyond the detection range $\unit{700}\pm\unit{20}{\nano\metre}$.

As may be observed from Fig.~\ref{fig-mm-tp}d, the luminescence spots on the metamaterials are not neatly localised unlike the continuous gold film (Fig.~\ref{fig-au-tp}b). Therefore, the nature of excitation cannot be inferred by simply studying the size of the luminescence spot with respect to the excitation spot. As an alternative, the power dependence of the metamaterial luminescence is studied (inset Fig.~\ref{fig-mm-tp}c). The log-log plot of luminescence intensity versus incident power for metamaterial C shows that the data points fit to a straight line with slope of $2.09\pm0.07$ implying a purely two-photon excitation process.

Besides the spectral shift of the emission peak, the introduction of localised surface plasmon resonances by nano-structuring of the continuous gold film leads to huge enhancement of the luminescence intensity. To study this, we move the illumination from the plain area of the gold film on to a metamaterial by scanning the sample stage in \unit{100}{\nano\metre} steps normal to the incident beam. At each point an image of the transmitted luminescence is recorded by the camera. This is repeated for the four metamaterials and with the different spectral filters for each of them. A peak power density of $\unit{40}{\giga\watt}/{\centi\metre}^2$ is used, meaning no luminescence can be detected from the unstructured gold. As the excitation spot approaches the edge of a metamaterial, the detected intensity of photoluminescence starts increasing. When the excitation spot moves away from the edge and well within the structured area the intensity of luminescence levels out. For a particular metamaterial, the enhancement of photoluminescence is maximum when the detection spectral range coincides with the emission peak of that metamaterial.

Besides the spectral shift of the emission peak, the introduction of localised surface plasmon resonances by nano-structuring of the continuous gold film leads to huge enhancement of the luminescence intensity. To study this, we move the illumination from the plain area of the gold film on to a metamaterial by scanning the sample stage in \unit{100}{\nano\metre} steps normal to the incident beam. At each point an image of the transmitted luminescence is recorded by the camera. This is repeated for the four metamaterials and with the different spectral filters for each of them. A peak power density of $\unit{40}{\giga\watt}/{\centi\metre}^2$ is used, meaning no luminescence can be detected from the unstructured gold. As the excitation spot approaches the edge of a metamaterial, the detected intensity of photoluminescence starts increasing. When the excitation spot moves away from the edge and well within the structured area the intensity of luminescence levels out. For a particular metamaterial, the enhancement of photoluminescence is maximum when the detection spectral range coincides with the emission peak of that metamaterial.

To quantify the luminescence enhancement from the metamaterials, we define a enhancement factor for a chosen metamaterial and a given detection spectral range. This is calculated from the scanning data by averaging the luminescence intensity on the metamaterial and normalising it by the average signal level from the unstructured gold film. As seen in Fig.~\ref{fig-bubble}a, for metamaterial A the luminescence from the continuous gold film is enhanced by at least 50 times within the spectral range $\unit{650}\pm\unit{20}{\nano\metre}$. The enhancement factor decreases outside this range, as we expect by studying its emission spectra in Fig.~\ref{fig-mm-tp}. The largest enhancement in photoluminescence is recorded due to the presence of metamaterial C over the detection spectral range $\unit{700}\pm\unit{20}{\nano\metre}$.

\section*{Discussion}

In the following sections we discuss the possible mechanism of luminescence enhancement due to nanostructuring of the gold film. Here we have observed that the presence of the metamaterials can alter the photoluminescence characteristics of a gold film in two ways: (1) red-shift the emission peak, and (2) enhance the intensity of the photo-luminescence over the whole detection spectral range. In general, local-field enhancement due to the presence of localised surface plasmon will have an impact on both the excitation and emission fields leading to overall intensity enhancement of the emission from the metamaterials. A strong correlation between emission spectra and linear absorption spectra suggests that the surface plasmons play an active role in enhancing the photoluminescence from the metamaterials.

The metamaterial resonances would play a role in enhancing the contribution of the excitation field as well as the emission field. Among the four different metamaterials used here, metamaterial C shows the strongest enhancement of luminescence, specially at the emission wavelength of \unit{700}{\nano\metre}. It is not a coincidence that this particular metamaterial is characterised by a transmission dip and an absorption peak at the wavelength of excitation (\unit{850}{\nano\metre}). The transmission dip is related to the closed plasmonic mode of the metamaterial which is very weakly coupled to the free space, and helps to design high quality factor absorption resonances on thin surfaces~\cite{Prosvirnin2002}. For a negative or complementary metamaterial, like the one used here, dark-modes of surface plasmons can be excited by incident polarisation parallel to the axis of symmetry of the metamolecule, as used here, which however vanishes for the orthogonal polarisation. The resulting non-symmetric resonance is well known as the Fano resonance~\cite{Prosvirnin2002,Papasimakis2009b,Lukyanchuk2010} which arises from the asymmetry in the metamolecule design. For metamaterial C, the wavelength of two-photon excitation occurs at the absorption peak of Fano-resonance resulting in huge field enhancement of the excitation beam. The localised fields at this resonance wavelength do not suffer from background radiation and act on the constituent gold atoms in the near-field regime to enhance their resultant photoluminescence~\cite{Stockman2008}.

\begin{figure}[ht]
\centering
\includegraphics[width=0.85\linewidth]{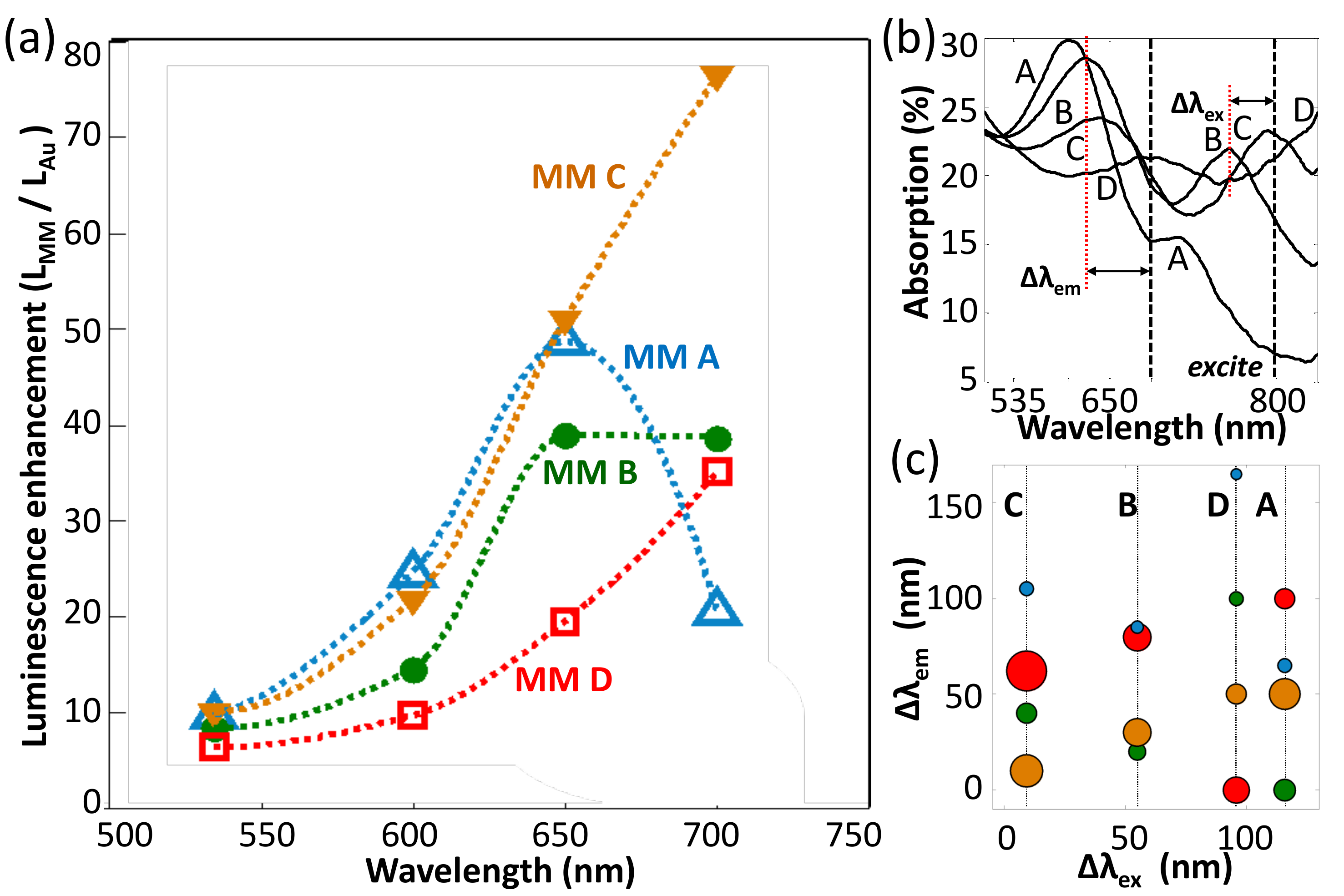}
\caption{Enhancement of luminescence. (a) Enhancement factor for different metamaterials. (b) Linear absorption of metamaterials and detuning parameters. (b) Enhancement factor as a function of emission and excitation detuning. Colours represent detection wavelength; Red: \unit{700}{\nano\metre}, Yellow: \unit{650}{\nano\metre}, Green: \unit{600}{\nano\metre}, and Blue: \unit{535}{\nano\metre}. Size of the bubbles represents strength of enhancement.}
\label{fig-bubble}
\end{figure}

To summarise the effect of the engineered plasmonic resonances on the photoluminescence of the metamaterials, we define two detuning parameters, one for emission and one for excitation, based on the position of the absorption resonances. Figure 4b shows the linear absorption spectra of the four different metamaterials when excited with white light linearly polarised parallel to the axis of symmetry of the metamolecules. These are calculated from the experimentally measured values of transmittance and reflectance of individual metamaterials (same as Fig. \ref{fig-mm-tp}b). For any particular metamaterial, there are two absorption peaks within the spectral range shown. By comparing the resonance peaks with known features of the resonances of split ring metamaterials, we can assign the peaks with a high degree of confidence. We first look at the absorption peak at longer wavelength: in separate measurements (not shown here), we observe that this longer wavelength resonance vanishes for the orthogonal incident polarisation and so we assign it to the high quality factor Fano resonance discussed in the last paragraph. This Fano-absorption peak can be excited only by the polarisation parallel to the axis of symmetry of the metamolecules\cite{Prosvirnin2002}.  The absorption peak at shorter wavelength is attributed to a bright mode of plasmonic resonance with non-zero dipole moment that couples well to the far-field and is easily excited with external illumination. Unlike the Fano-resonance peak, this dipole-absorption peak is also present for the incident polarisation that is perpendicular to the axis of symmetry of the metamaterials, but it appears as a low quality factor resonance with much wider linewidth. Based on these two absorption peaks, we define the detuning parameters $\Delta\lambda_{em}$ and $\Delta\lambda_{ex}$ for each metamaterial. $\Delta\lambda_{em}$ gives a measure of the difference in position of the (shorter wavelength) dipole-absorption peak from the corresponding emission wavelength and $\Delta\lambda_{ex}$ measures the difference in the position of the (longer wavelength) Fano-absorption peak from the excitation wavelength.

Figure~\ref{fig-bubble}c presents a comprehensive plot summarising the effects of these detuning parameters on the luminescence enhancement from the metamaterials. Each vertical column represents each metamaterial and different colours of the bubbles represent different emission wavelength while the relative size of the bubbles give a measure of the luminescence enhancement. The metamaterial C which has the least $\Delta\lambda_{ex}$ among all the metamaterials shows maximum enhancement at \unit{700}{\nano\metre}, though $\Delta\lambda_{em}$ is minimum for \unit{650}{\nano\metre}. This may be explained by Stoke's shift which would only red-shift the emission peak with respect to the absorption resonance. For the metamaterial B, where $\Delta\lambda_{ex}$ is slightly more than \unit{50}{\nano\metre}, the enhancement is similar for \unit{650}{\nano\metre} and \unit{700}{\nano\metre}, but neither as large as for the metamaterial C at \unit{700}{\nano\metre}. In fact, among all the recorded enhancement, the metamaterial C at \unit{700}{\nano\metre} provides the highest. It is interesting to note that for the metamaterial D the luminescence enhancement at \unit{700}{\nano\metre} is weaker than the metamaterial C even though $\Delta\lambda_{em}=0$ for the former. Observing the overall trend of luminescence enhancement presented in the Fig.~\ref{fig-bubble}c it may be inferred that it is crucial to provide excitation at the closed-mode of plasmonic resonance of the metamaterial.

Finally, for the metamaterials A, B, and D whose Fano-absorption peak is far from the excitation wavelength, an interesting theory on luminescence enhancement may be considered as given by Dulkeith \emph{et al.}~\cite{Dulkeith2004}. While studying gold nanoparticles Dulkeith \emph{et al.} observed that the enhancement in photoluminescence cannot be simply explained by the local-field enhancement due to surface plasmons. They proposed that the dominant contribution to the photoluminescence originated from radiative relaxation of particle-plasmons which were created by hot carriers~\cite{Shahbazyan1998}. In Dulkeith's system, only interband transitions were excited and not plasmonic resonances. Hence no local-field enhancement was expected from the excitation process. This is similar to the case of all the metamaterials except C. The emission process was, however, influenced by plasmonic resonance. Radiative decay of surface plasmons has been reported by others~\cite{Basak2006}. This may be one of the processes explaining the enhanced luminescence from our planar metamaterials and is worth investigating in future work.

In conclusion, we have experimentally demonstrated that the two-photon excited photoluminescence of gold can be controlled by nanopatterning an ultrathin gold film into metamaterials with designed absorption resonances. In the presence of the metamaterials, the emission peak of the unstructured gold can be tuned to higher spectral regime and the intensity of the photoluminescence can be significantly enhanced. The shift of the emission peak is associated with the bright mode of absorption resonance of the metamaterials. An enhancement of at least 76 times is achieved when resonant two-photon excitation is tuned to the dark-mode plasmonic absorption of the metamaterial.


\begin{thebibliography}{10}
\expandafter\ifx\csname url\endcsname\relax
  \def\url#1{\texttt{#1}}\fi
\expandafter\ifx\csname urlprefix\endcsname\relax\def\urlprefix{URL }\fi
\providecommand{\bibinfo}[2]{#2}
\providecommand{\eprint}[2][]{\url{#2}}

\bibitem{Herschel1845}
\bibinfo{author}{Herschel, J. F.~W.}
\newblock \bibinfo{title}{On a case of superficial colour presented by a
  homogeneous liquid internally colourless}.
\newblock \emph{\bibinfo{journal}{Philosophical Transactions of the Royal
  Society of London}} \textbf{\bibinfo{volume}{135}}, \bibinfo{pages}{143--145}
  (\bibinfo{year}{1845}).

\bibitem{Lakowicz2006}
\bibinfo{author}{Lakowicz, J.~R.}
\newblock \emph{\bibinfo{title}{Principles of fluorescence spectroscopy}}
  (\bibinfo{publisher}{Springer}, \bibinfo{year}{2006}).

\bibitem{Wood1928}
\bibinfo{author}{Wood, R.} \& \bibinfo{author}{Gaviola, E.}
\newblock \bibinfo{title}{The power relation of the intensities of the lines in
  the optical excitation of mercury}.
\newblock \emph{\bibinfo{journal}{Philosophical Magazine}}
  \textbf{\bibinfo{volume}{6}}, \bibinfo{pages}{352 -- 356}
  (\bibinfo{year}{1928}).

\bibitem{Goeppert-Mayer1931}
\bibinfo{author}{Goeppert-Mayer, M.}
\newblock \emph{\bibinfo{title}{Uber Elementarakte mit zwei Quantensprungen}}.
\newblock Ph.D. thesis, \bibinfo{school}{University of Gottingen, Germany}
  (\bibinfo{year}{1931}).

\bibitem{Kaiser1961}
\bibinfo{author}{Kaiser, W.} \& \bibinfo{author}{Garrett, C. G.~B.}
\newblock \bibinfo{title}{Two-photon excitation in
  ca$\textrm{F}^{2}:\textrm{Eu}^{2+}$}.
\newblock \emph{\bibinfo{journal}{Phys. Rev. Lett.}}
  \textbf{\bibinfo{volume}{7}}, \bibinfo{pages}{229--231}
  (\bibinfo{year}{1961}).

\bibitem{Maiman1960}
\bibinfo{author}{Maiman, T.~H.}
\newblock \bibinfo{title}{Stimulated optical radiation in ruby}.
\newblock \emph{\bibinfo{journal}{Nature}} \textbf{\bibinfo{volume}{187}}
  (\bibinfo{year}{1960}).

\bibitem{Mooradian1969}
\bibinfo{author}{Mooradian, A.}
\newblock \bibinfo{title}{{Photoluminescence of Metals}}.
\newblock \emph{\bibinfo{journal}{Physical Review Letters}}
  \textbf{\bibinfo{volume}{22}}, \bibinfo{pages}{5--7} (\bibinfo{year}{1969}).

\bibitem{Boyd1986}
\bibinfo{author}{Boyd, G.~T.}, \bibinfo{author}{Yu, Z.~H.} \&
  \bibinfo{author}{Shen, Y.~R.}
\newblock \bibinfo{title}{Photoinduced luminescence from the noble metals and
  its enhancement on roughened surfaces}.
\newblock \emph{\bibinfo{journal}{Phys. Rev. B}} \textbf{\bibinfo{volume}{33}},
  \bibinfo{pages}{7923--7936} (\bibinfo{year}{1986}).

\bibitem{Beversluis2003}
\bibinfo{author}{Beversluis, M.}, \bibinfo{author}{Bouhelier, A.} \&
  \bibinfo{author}{Novotny, L.}
\newblock \bibinfo{title}{{Continuum generation from single gold nanostructures
  through near-field mediated intraband transitions}}.
\newblock \emph{\bibinfo{journal}{Physical Review B}}
  \textbf{\bibinfo{volume}{68}}, \bibinfo{pages}{115433}
  (\bibinfo{year}{2003}).

\bibitem{Bouhelier2003}
\bibinfo{author}{Bouhelier, A.}, \bibinfo{author}{Beversluis, M.~R.} \&
  \bibinfo{author}{Novotny, L.}
\newblock \bibinfo{title}{{Characterization of nanoplasmonic structures by
  locally excited photoluminescence}}.
\newblock \emph{\bibinfo{journal}{Applied Physics Letters}}
  \textbf{\bibinfo{volume}{83}}, \bibinfo{pages}{5041} (\bibinfo{year}{2003}).

\bibitem{Link2003}
\bibinfo{author}{Link, S.} \& \bibinfo{author}{El-Sayed, M.~A.}
\newblock \bibinfo{title}{Optical properties and ultrafast dynamics ofmetallic
  nanocrystals}.
\newblock \emph{\bibinfo{journal}{Annu. Rev. Phys. Chem}}
  \textbf{\bibinfo{volume}{54}}, \bibinfo{pages}{331–66}
  (\bibinfo{year}{2003}).

\bibitem{Imura2006}
\bibinfo{author}{Imura, K.}, \bibinfo{author}{Nagahara, T.} \&
  \bibinfo{author}{Okamoto, H.}
\newblock \bibinfo{title}{{Photoluminescence from gold nanoplates induced by
  near-field two-photon absorption}}.
\newblock \emph{\bibinfo{journal}{Applied Physics Letters}}
  \textbf{\bibinfo{volume}{88}}, \bibinfo{pages}{023104}
  (\bibinfo{year}{2006}).

\bibitem{Wang2009}
\bibinfo{author}{Wang, D.-S.}, \bibinfo{author}{Hsu, F.-Y.} \&
  \bibinfo{author}{Lin, C.-W.}
\newblock \bibinfo{title}{{Surface plasmon effects on two photon luminescence
  of gold nanorods.}}
\newblock \emph{\bibinfo{journal}{Optics Express}}
  \textbf{\bibinfo{volume}{17}}, \bibinfo{pages}{11350--9}
  (\bibinfo{year}{2009}).

\bibitem{Beermann2012}
\bibinfo{author}{Beermann, J.} \emph{et~al.}
\newblock \bibinfo{title}{{Polarization-resolved two-photon luminescence
  microscopy of V-groove arrays.}}
\newblock \emph{\bibinfo{journal}{Optics Express}}
  \textbf{\bibinfo{volume}{20}}, \bibinfo{pages}{654--62}
  (\bibinfo{year}{2012}).

\bibitem{Jiang2013}
\bibinfo{author}{Jiang, X.-F.} \emph{et~al.}
\newblock \bibinfo{title}{{Excitation Nature of Two-Photon Photoluminescence of
  Gold Nanorods and Coupled Gold Nanoparticles Studied by Two-Pulse Emission
  Modulation Spectroscopy}}.
\newblock \emph{\bibinfo{journal}{The Journal of Physical Chemistry Letters}}
  \textbf{\bibinfo{volume}{4}}, \bibinfo{pages}{1634--1638}
  (\bibinfo{year}{2013}).

\bibitem{Tao2013}
\bibinfo{author}{Tao, W.}, \bibinfo{author}{Bao, H.} \& \bibinfo{author}{Gu,
  M.}
\newblock \bibinfo{title}{{Two-photon-excited photoluminescence and heating of
  gold nanorods through absorption of supercontinuum light}}.
\newblock \emph{\bibinfo{journal}{Applied Physics B}}
  \textbf{\bibinfo{volume}{112}}, \bibinfo{pages}{153--158}
  (\bibinfo{year}{2013}).

\bibitem{Wang2013}
\bibinfo{author}{Wang, T.}, \bibinfo{author}{Halaney, D.}, \bibinfo{author}{Ho,
  D.}, \bibinfo{author}{Feldman, M.~D.} \& \bibinfo{author}{Milner, T.~E.}
\newblock \bibinfo{title}{{Two-photon luminescence properties of gold
  nanorods}}.
\newblock \emph{\bibinfo{journal}{Biomedical Optics Express}}
  \textbf{\bibinfo{volume}{4}}, \bibinfo{pages}{584--595}
  (\bibinfo{year}{2013}).

\bibitem{Zhao2012}
\bibinfo{author}{Zhao, T.} \emph{et~al.}
\newblock \bibinfo{title}{Gold nanorods as agents for two-photon photodynamic
  therapy}.
\newblock \emph{\bibinfo{journal}{Nanoscale}} \bibinfo{pages}{7712--7719}
  (\bibinfo{year}{2012}).

\bibitem{Rivera2012}
\bibinfo{author}{Rivera, V. A.~G.}, \bibinfo{author}{Ferri, F.~A.} \&
  \bibinfo{author}{Jr, E.~M.}
\newblock \emph{\bibinfo{title}{{Plasmonics - Principles and Applications}}}
  (\bibinfo{publisher}{InTech}, \bibinfo{year}{2012}).

\bibitem{Nahm2013}
\bibinfo{author}{Nahm, C.} \emph{et~al.}
\newblock \bibinfo{title}{{Photoluminescence Enhancement by Surface-Plasmon
  Resonance: Recombination-Rate Theory and Experiments}}.
\newblock \emph{\bibinfo{journal}{Applied Physics Express}}
  \textbf{\bibinfo{volume}{6}}, \bibinfo{pages}{052001} (\bibinfo{year}{2013}).

\bibitem{Song2012}
\bibinfo{author}{Song, M.} \emph{et~al.}
\newblock \bibinfo{title}{Polarization properties of surface plasmon enhanced
  photoluminescence from a single ag nanowire}.
\newblock \emph{\bibinfo{journal}{Optics Express}}
  \textbf{\bibinfo{volume}{20}}, \bibinfo{pages}{22290--22297}
  (\bibinfo{year}{2012}).

\bibitem{Su2010}
\bibinfo{author}{Su, Y.~H.} \emph{et~al.}
\newblock \bibinfo{title}{Influence of surface plasmon resonance on the
  emission intermittency of photoluminescence from gold nano-sea-urchins}.
\newblock \emph{\bibinfo{journal}{Nanoscale}} \textbf{\bibinfo{volume}{2}},
  \bibinfo{pages}{2639--2646} (\bibinfo{year}{2010}).

\bibitem{Rockstuhl2006}
\bibinfo{author}{Rockstuhl, C.} \emph{et~al.}
\newblock \bibinfo{title}{On the reinterpretation of resonances in
  split-ring-resonators at normal incidence}.
\newblock \emph{\bibinfo{journal}{Optics Express}}
  \textbf{\bibinfo{volume}{14}}, \bibinfo{pages}{8827--8836}
  (\bibinfo{year}{2006}).

\bibitem{Tanaka2010}
\bibinfo{author}{Tanaka, K.}, \bibinfo{author}{Plum, E.}, \bibinfo{author}{Ou,
  J.~Y.}, \bibinfo{author}{Uchino, T.} \& \bibinfo{author}{Zheludev, N.~I.}
\newblock \bibinfo{title}{{Multifold Enhancement of Quantum Dot Luminescence in
  Plasmonic Metamaterials}}.
\newblock \emph{\bibinfo{journal}{Physical Review Letters}}
  \textbf{\bibinfo{volume}{105}}, \bibinfo{pages}{227403}
  (\bibinfo{year}{2010}).

\bibitem{Nikolaenko2012}
\bibinfo{author}{Nikolaenko, A.~E.} \emph{et~al.}
\newblock \bibinfo{title}{Nonlinear graphene metamaterial}.
\newblock \emph{\bibinfo{journal}{Applied Physics Letters}}
  \textbf{\bibinfo{volume}{100}}, \bibinfo{pages}{--} (\bibinfo{year}{2012}).

\bibitem{Ren2011}
\bibinfo{author}{Ren, M.} \emph{et~al.}
\newblock \bibinfo{title}{{Nanostructured plasmonic medium for terahertz
  bandwidth all-optical switching.}}
\newblock \emph{\bibinfo{journal}{Advanced materials (Deerfield Beach, Fla.)}}
  \textbf{\bibinfo{volume}{23}}, \bibinfo{pages}{5540--4}
  (\bibinfo{year}{2011}).

\bibitem{Nikolaenko2010}
\bibinfo{author}{Nikolaenko, A.~E.} \emph{et~al.}
\newblock \bibinfo{title}{Carbon nanotubes in a photonic metamaterial}.
\newblock \emph{\bibinfo{journal}{Phys. Rev. Lett.}}
  \textbf{\bibinfo{volume}{104}}, \bibinfo{pages}{153902}
  (\bibinfo{year}{2010}).

\bibitem{Adamo2012}
\bibinfo{author}{Adamo, G.} \emph{et~al.}
\newblock \bibinfo{title}{Electron-beam-driven collective-mode metamaterial
  light source}.
\newblock \emph{\bibinfo{journal}{Phys. Rev. Lett.}}
  \textbf{\bibinfo{volume}{109}}, \bibinfo{pages}{217401}
  (\bibinfo{year}{2012}).

\bibitem{Plum2009}
\bibinfo{author}{Plum, E.}, \bibinfo{author}{Fedotov, V.~A.},
  \bibinfo{author}{Kuo, P.}, \bibinfo{author}{Tsai, D.~P.} \&
  \bibinfo{author}{Zheludev, N.~I.}
\newblock \bibinfo{title}{Towards the lasing spaser: controlling metamaterial
  optical response with semiconductor quantum dots}.
\newblock \emph{\bibinfo{journal}{Opt. Express}} \textbf{\bibinfo{volume}{17}},
  \bibinfo{pages}{8548--8551} (\bibinfo{year}{2009}).

\bibitem{Adamo2013}
\bibinfo{author}{Adamo, G.} \emph{et~al.}
\newblock \bibinfo{title}{Tunable light emission in reconfigurable plasmonic
  metamaterials}.
\newblock In \emph{\bibinfo{booktitle}{Lasers and Electro-Optics Europe (CLEO
  EUROPE/IQEC), 2013 Conference on and International Quantum Electronics
  Conference}}, \bibinfo{pages}{1--1} (\bibinfo{year}{2013}).

\bibitem{Coherent2012}
\bibinfo{author}{Coherent}.
\newblock \bibinfo{title}{Chameleon vision s – the next generation of one-box
  ti:sapphire lasers} (\bibinfo{year}{2012}).

\bibitem{Stokes1852}
\bibinfo{author}{{Stokes}, G.~G.}
\newblock \bibinfo{title}{{On the Change of Refrangibility of Light}}.
\newblock \emph{\bibinfo{journal}{Royal Society of London Philosophical
  Transactions Series I}} \textbf{\bibinfo{volume}{142}},
  \bibinfo{pages}{463--562} (\bibinfo{year}{1852}).

\bibitem{Prosvirnin2002}
\bibinfo{author}{Prosvirnin, S.} \& \bibinfo{author}{Zouhdi, S.}
\newblock \bibinfo{title}{Resonances of closed modes in thin arrays of complex
  particles}.
\newblock In \bibinfo{editor}{Zouhdi, S.}, \bibinfo{editor}{Sihvola, A.} \&
  \bibinfo{editor}{Arsalane, M.} (eds.) \emph{\bibinfo{booktitle}{Advances in
  Electromagnetics of Complex Media and Metamaterials}},
  vol.~\bibinfo{volume}{89} of \emph{\bibinfo{series}{NATO Science Series}},
  \bibinfo{pages}{281--290} (\bibinfo{publisher}{Springer Netherlands},
  \bibinfo{year}{2002}).

\bibitem{Papasimakis2009b}
\bibinfo{author}{Papasimakis, N.} \& \bibinfo{author}{Zheludev, N.~I.}
\newblock \bibinfo{title}{Metamaterial-induced transparency:sharp fano
  resonances and slow light}.
\newblock \emph{\bibinfo{journal}{Opt. Photon. News}}
  \textbf{\bibinfo{volume}{20}}, \bibinfo{pages}{22--27}
  (\bibinfo{year}{2009}).

\bibitem{Lukyanchuk2010}
\bibinfo{author}{Luk'yanchuk, B.} \emph{et~al.}
\newblock \bibinfo{title}{{The Fano resonance in plasmonic nanostructures and
  metamaterials.}}
\newblock \emph{\bibinfo{journal}{Nature Materials}}
  \textbf{\bibinfo{volume}{9}}, \bibinfo{pages}{707--15}
  (\bibinfo{year}{2010}).

\bibitem{Stockman2008}
\bibinfo{author}{Stockman, M.~I.}
\newblock \bibinfo{title}{Spasers explained}.
\newblock \emph{\bibinfo{journal}{Nature Photonics}}
  \textbf{\bibinfo{volume}{2}}, \bibinfo{pages}{327--329}
  (\bibinfo{year}{2008}).

\bibitem{Dulkeith2004}
\bibinfo{author}{Dulkeith, E.} \emph{et~al.}
\newblock \bibinfo{title}{Plasmon emission in photoexcited gold nanoparticles}.
\newblock \emph{\bibinfo{journal}{Phys. Rev. B}} \textbf{\bibinfo{volume}{70}},
  \bibinfo{pages}{205424} (\bibinfo{year}{2004}).

\bibitem{Shahbazyan1998}
\bibinfo{author}{Shahbazyan, T.~V.}, \bibinfo{author}{Perakis, I.~E.} \&
  \bibinfo{author}{Bigot, J.-Y.}
\newblock \bibinfo{title}{Size-dependent surface plasmon dynamics in metal
  nanoparticles}.
\newblock \emph{\bibinfo{journal}{Phys. Rev. Lett.}}
  \textbf{\bibinfo{volume}{81}}, \bibinfo{pages}{3120--3123}
  (\bibinfo{year}{1998}).

\bibitem{Basak2006}
\bibinfo{author}{Basak, D.}, \bibinfo{author}{Karan, S.} \&
  \bibinfo{author}{Mallik, B.}
\newblock \bibinfo{title}{Size selective photoluminescence in poly(methyl
  methacrylate) thin solid films with dispersed silver nanoparticles
  synthesized by a novel method}.
\newblock \emph{\bibinfo{journal}{Chemical Physics Letters}}
  \textbf{\bibinfo{volume}{420}}, \bibinfo{pages}{115 -- 119}
  (\bibinfo{year}{2006}).

\end{thebibliography}

\section*{Acknowledgements}

This work was funded by the United Kingdom Engineering and Physical Sciences Research Council, grant number EP/F040644/1, the University of Southampton Enterprise Fund, and the Advanced Optics in Engineering Programme, A*STAR, Singapore, grant number 122-360-0009.

\section*{Author contributions statement}

T.R. conducted the experiments and performed the data analysis, T.R. and E.R. designed the experimental setup, N.Z initiated and supervised the work. All authors discussed and interpreted the results extensively and reviewed the manuscript.

\section*{Additional information}

Competing financial interests The authors declare no competing financial interests.
The data from this paper can be obtained from the University of Southampton ePrints research repository: URL http://eprints.soton.ac.uk/id/eprint/383676

%
%

\end{document}